\documentclass[aps,pra,preprint,superscriptaddress,longbibliography]{revtex4-2}

\usepackage{amsmath}
\usepackage{amsthm}
\usepackage{amsfonts}
\usepackage{amssymb}
\usepackage{xcolor,graphicx}
\usepackage{tikz}
\usepackage{color}
\usepackage[pdftex,hidelinks]{hyperref} 
\usepackage{longtable}
\usepackage{array}
\usepackage{dsfont}
\usepackage{bm}
\definecolor{r2revision}{RGB}{0,0,0}

\begin{document}
	
	\title{Self-similar rigid rotation in anisotropic squeezed Laguerre--Gaussian beams}
	
	\author{D. Aguirre-Olivas}
	\email[e-mail: ]{dilia.aguirre@inaoe.mx}
	\affiliation{Instituto Nacional de Astrof\'isica, \'Optica y Electr\'onica, Tonantzintla, Puebla 72840, Mexico.}
	
	\author{G. Mellado-Villase\~nor}
	\email[e-mail: ]{gmelladov@inaoe.mx}
	\affiliation{Instituto Nacional de Astrof\'isica, \'Optica y Electr\'onica, Tonantzintla, Puebla 72840, Mexico.}
	
	\author{B.~M. Rodr\'iguez-Lara}
	\email[e-mail: ]{blas.rodriguez@gmail.com}
	\affiliation{Universidad Polit\'ecnica Metropolitana de Hidalgo, Tolcayuca, Hidalgo 43860, Mexico.}
	
	\date{\today}
	
	\begin{abstract}
		We introduce anisotropic squeezed Laguerre--Gaussian beams as an analytically complete family of structured paraxial scalar beams with Gouy-phase-controlled \textcolor{r2revision}{self-similar rigid rotation}.
		Our construction organizes the Laguerre--Gaussian transverse modal Hilbert space into circular modal subspaces carrying parity-resolved single-mode $\mathrm{su}(1,1)$ representations.
		The resulting optical Bargmann bases support squeezed states whose lowest-weight forms factor into Gaussian beam phases, parity-dependent polynomials, and elliptically anisotropic Gaussian envelopes.
		Excited seeds are polynomial deformations of the same squeezed envelope, extending the analytic construction to the full family.
		The Gouy-dressed squeezing parameter sets the semi-axis widths, aspect ratio, handedness, and orientation, while propagation rotates the profile by the accumulated Gouy phase without changing the scaled transverse shape.
		We generate these fields with synthetic phase holograms and experimentally confirm their self-similar rigid rotation.
		Our scalar-beam family provides a tunable anisotropic structured-light platform whose transverse orientation encodes propagation distance and may support axial localization, range finding, optical manipulation, and modal diagnostics.
	\end{abstract}
	
	\maketitle
	\newpage

	
	\section{Introduction} 
	\label{sec:Sec1}
	
	Laguerre--Gaussian beams (LGBs) form a complete orthonormal basis for structured paraxial scalar fields.
	Their radial and azimuthal degrees of freedom carry complementary $\mathrm{su}(1,1)$~\cite{Karimi2014p063813,Plick2015p063841} and $\mathrm{su}(2)$~\cite{Allen1992p8185,Padgett1999p430} algebraic structures.
	Circular ladder operators generate the full LGB basis from the fundamental Gaussian and organize both degrees of freedom through circular occupation numbers~\cite{Nienhuis1993p656}; their minimum gives the radial mode number, while their imbalance gives the azimuthal mode number.
	Under free-space propagation, all LGBs share the same Gaussian beam scaling and curvature factors, while each mode accumulates a Gouy phase proportional to its total modal number~\cite{Siegman1986}.
	The modal structure and Gouy phase accumulation make LGBs a natural platform for engineering structured paraxial scalar beams with prescribed transverse evolution.
	
	Controlled transverse rotation during free-space propagation appears in spiral beams~\cite{Abramochkin1993p336}, twisted Gaussian Schell-model beams~\cite{Simon1993p95,Simon1993p2008}, astigmatic Gaussian modes~\cite{Visser2004p013809,Simon1999p2465}, and self-rotating beams~\cite{Niu2022p5465,Niu2023p12150,Han2024p130402,Jiang2025p25007}.
	In these examples, rotation is generated by a field or phase ansatz, partial coherence, or a fixed astigmatic geometry.
	\textcolor{black}{Bessel-beam superpositions support rotating vortex arrays~\cite{Lusk2023p023509} and controlled vortex interactions~\cite{Voitiv2024p016135}, while balanced vortex--antivortex couplings can preserve lattice structures during free-space propagation~\cite{Lin2024p6178}.}
	Gouy phase accumulation provides a modal route for controlled transverse rotation.
	It produces observable rotation of asymmetric vortex features in modified LGBs~\cite{Hamazaki2006p8382}, drives off-axis vortex motion in LGB superpositions~\cite{Baumann2009p9818}, governs propagation variations and revivals in structured vector light~\cite{Zhong2021p053520,PinheirodaSilva2020p033902}, controls rigid and nonrigid transverse rotation in products of displaced LGBs through their net orbital angular momentum~\cite{MelladoVillasenor2026p023514}, and encodes axial position through rotating point-spread functions in three-dimensional localization microscopy~\cite{Greengard2006p181,Pavani2008p3484}.
	\textcolor{black}{Coaxial LGB superpositions also exhibit propagation-dependent rotation~\cite{Huang2016p132}.}
	\textcolor{r2revision}{LGB superpositions rotate self-similarly when their modal-order differences are proportional to their azimuthal-index differences~\cite{Schechner1996pR50}.}
	
	\textcolor{r2revision}{We combine this modal rotation condition with algebraic squeezing in transverse mode spaces.}
	Algebraic methods have produced classical structured light analogues of quantum coherent and squeezed states.
	Coherent state constructions in LGB circular mode subspaces, and generalized constructions in Hermite--Gaussian and LGB bases, reproduce displaced, rotated, and squeezed coherent-state structures in classical paraxial fields~\cite{MoralesRodriguez2024p1489,TorresLeal2024p063716,AguirreOlivas2025p1121,MoralesRodriguez2024p033523}.
	These approaches have recently been extended to vector structured light through total angular momentum coherent state fields based on the coupling of polarization and spatial degrees of freedom~\cite{AguirreOlivas2026p3221}.
	In the Hermite--Gaussian basis, $\mathrm{su}(1,1)$ squeezing connects structured light squeezing to tight focusing and spatial metrology~\cite{Wang2024p297}.
	In the LGB basis, the radial-index $\mathrm{su}(1,1)$ \textcolor{black}{Lie algebra} realization maps single-mode squeezed number states onto orthonormal LGB superpositions and establishes classical analogues of continuous-variable correlations~\cite{RodriguezLara2026p015032}.
	That radial realization raises both circular occupation numbers at fixed azimuthal charge, producing squeezed LGBs with self-similar propagation but no rigid rotation.
	\textcolor{r2revision}{Our circular-mode squeezing keeps one occupation fixed while varying the other, enforcing self-similar rigid rotation and organizing these fields into parity-resolved families.}
	
	Here, we introduce anisotropic squeezed Laguerre--Gaussian beams \textcolor{r2revision}{that undergo self-similar rigid rotation} set by the accumulated Gouy phase.
	We organize the LGB transverse modal Hilbert space into circular modal subspaces carrying single-mode $\mathrm{su}(1,1)$ representations, Sec.~\ref{sec:Sec2}.
	In each subspace, one circular occupation number labels the subspace, while the other defines a parity-resolved optical Bargmann ladder.
	Applying the corresponding squeezing operators to these optical bases yields a closed lowest-weight seed form and a recursion formula for excited seeds, Sec.~\ref{sec:Sec3}.
	The lowest-weight form factors into Gaussian beam curvature and Gouy phases, a parity-dependent polynomial, and an elliptically anisotropic Gaussian envelope.
	The Gouy phase dresses the squeezing parameter in the polynomial factor and Gaussian envelope, preserving the scaled semi-axis widths and aspect ratio while rotating the principal axis by the accumulated Gouy phase, Sec.~\ref{sec:Sec4}.
	In Sec.~\ref{sec:Sec5}, we describe the experimental implementation through synthetic phase holograms.
	We derive the \textcolor{r2revision}{self-similar rigid rotation law}, experimentally confirm \textcolor{r2revision}{its predictions}, and close in Sec.~\ref{sec:Sec6} by discussing applications in imaging, axial localization, optical manipulation, and modal diagnostics.
	
	\section{LGB circular-mode subspaces}
	\label{sec:Sec2}
	
	We work with orthonormal Laguerre--Gaussian beams (LGBs) \cite{AguirreOlivas2025p1121,MelladoVillasenor2026p023514},
	\begin{align}
		\begin{aligned}
			& \Psi_{p,\ell}(\bm{\rho},z) = \frac{\sqrt{2}}{w(z)} e^{ - i \frac{k \rho^{2}}{2 R(z)}} e^{i \left( N + 1 \right) \varphi(z)} \psi_{p,\ell}\left( q_{\rho}, q_{\phi} \right), \\ 
			& \int_{0}^{2\pi} d\phi \int_{0}^{\infty} d\rho \, \rho \, \Psi_{p^{\prime}, \ell^{\prime}}^{\ast}(\bm{\rho},z) \Psi_{p, \ell}(\bm{\rho},z) =~ \delta_{p, p^{\prime}}\delta_{\ell, \ell^{\prime}},
		\end{aligned}
	\end{align}
	where $\bm{\rho} = \left( \rho, \phi \right)$ is the transverse plane coordinate vector, $k = 2 \pi / \lambda$ is the wavenumber, $w(z) = w_{0} \sqrt{1 + q_{z}^{2}}$ is the beam width, $R(z) = z \left( 1 + q_{z}^{-2} \right)$ is the spherical phase curvature radius, and $\varphi(z) = \arctan q_{z}$ is the Gouy phase, with beam waist $w_{0}$, Rayleigh range $z_{R} = k w_{0}^{2} / 2$, and dimensionless propagation coordinate $q_{z} = z / z_{R}$.
	We use the dimensionless transverse coordinates $q_{\rho} \equiv q_{\rho}(z) = \sqrt{2} \, \rho / w(z)$ and $q_{\phi} = \phi$ in the normalized transverse profile,
	\begin{align}
		\psi_{p,\ell}(q_{\rho}, q_{\phi}) =&~ (-1)^{p} \sqrt{\frac{p!}{\pi \left( p + \lvert \ell \rvert \right)!}} \, q_{\rho}^{\lvert \ell \rvert} e^{- \frac{1}{2} q_{\rho}^{2}} L_{p}^{\lvert \ell \rvert}(q_{\rho}^{2}) e^{i \ell q_{\phi}},
	\end{align}
	which carries radial index $p = 0, 1, 2, \ldots$ and azimuthal index $\ell = 0, \pm1, \pm2, \ldots$, with total transverse modal number, or mode order,
	\begin{align}
		N =&~ 2 p + \lvert \ell \rvert.
	\end{align}
	
	\textcolor{black}{Classical operator methods provide a framework for describing beam propagation~\cite{Stoler1981p334,vanEnk1992p147}.}
	The orthonormal LGB family defines a modal Hilbert space spanned by basis vectors labeled by the discrete mode indices $(p,\ell)$.
	We use Dirac notation, $\lvert p,\ell\rangle$, for the modal basis vector associated with the scalar beam $\Psi_{p,\ell}(\bm{\rho},z)$ through the projection
	\begin{align}
		\langle \bm{\rho},z \vert p,\ell\rangle =&~ \Psi_{p,\ell}(\bm{\rho},z).
	\end{align}
	On this modal Hilbert space, we use the dimensionless Cartesian coordinates $q_{x}=q_{\rho}\cos q_{\phi}$ and $q_{y}=q_{\rho}\sin q_{\phi}$, the canonical product and differential operators $\hat{q}_{j}=q_{j}$ and $\hat{p}_{j}=-i\partial_{q_{j}}$, with $j=x,y$ and $\left[ \hat{q}_{j},\hat{p}_{k} \right]=i\delta_{j,k}$, and the Cartesian ladder operators $\hat{a}_{j}=\left( \hat{q}_{j}+i\hat{p}_{j} \right)/\sqrt{2}$.
	We define the circular ladder and number operators,
	\begin{align}
		\begin{aligned}
			\hat{a}_{\pm} =&~ \frac{1}{\sqrt{2}} \left( \hat{a}_{x} \mp i \hat{a}_{y} \right), \\
			\hat{n}_{\pm} =&~ \hat{a}_{\pm}^{\dagger}\hat{a}_{\pm},
		\end{aligned}
	\end{align}
	that are diagonal in the LGB basis,
	\begin{align}
		\begin{aligned}
			\hat{n}_{\pm}\lvert p,\ell\rangle =&~ n_{\pm}\lvert p,\ell\rangle, \\
			n_{\pm} =&~ p + \frac{1}{2}\left( \lvert \ell \rvert \pm \ell \right),
		\end{aligned}
	\end{align}
	with total transverse modal number operator
	\begin{align}
		\hat{N} =&~ \hat{n}_{+} + \hat{n}_{-}.
	\end{align}
	The paired circular ladders shift the radial index at fixed azimuthal index \cite{Karimi2014p063813,Plick2015p063841},
	\begin{align}
		\begin{aligned}
			\hat{a}_{+}^{\dagger}\hat{a}_{-}^{\dagger} \lvert p, \ell\rangle
			=&~ \sqrt{\left( n_{+} + 1 \right)\left( n_{-} + 1 \right)} \, \lvert p + 1, \ell\rangle, \\
			\hat{a}_{+}\hat{a}_{-} \lvert p, \ell\rangle
			=&~ \sqrt{n_{+} n_{-}} \, \lvert p - 1, \ell\rangle,
		\end{aligned}
	\end{align}
	while the single circular ladders shift one circular occupation at a time,
	\begin{align}
		\begin{aligned}
			\hat{a}_{+}^{\dagger} \lvert p, \ell\rangle
			=&~
			\begin{cases}
				\sqrt{p + \ell + 1} \, \lvert p, \ell + 1\rangle, & \ell \geq 0, \\
				\sqrt{p + 1} \, \lvert p + 1, \ell + 1\rangle, & \ell < 0,
			\end{cases} \\
			\hat{a}_{+} \lvert p, \ell\rangle
			=&~
			\begin{cases}
				\sqrt{p + \ell} \, \lvert p, \ell - 1 \rangle, & \ell > 0, \\
				\sqrt{p} \, \lvert p - 1, \ell - 1\rangle, & \ell \leq 0,
			\end{cases} \\
			\hat{a}_{-}^{\dagger} \lvert p, \ell\rangle
			=&~
			\begin{cases}
				\sqrt{p - \ell + 1} \, \lvert p, \ell - 1 \rangle, & \ell \leq 0, \\
				\sqrt{p + 1} \, \lvert p + 1, \ell - 1 \rangle, & \ell > 0,
			\end{cases} \\
			\hat{a}_{-} \lvert p, \ell \rangle
			=&~
			\begin{cases}
				\sqrt{p-\ell} \, \lvert p, \ell + 1 \rangle, & \ell < 0, \\
				\sqrt{p} \, \lvert p - 1, \ell + 1 \rangle, & \ell \geq 0.
			\end{cases}
		\end{aligned}
	\end{align}
	These relations follow from the classical differential operator structure of the LGB modal space; hats are profile operators, and kets are modal basis vectors, with no quantum interpretation implied.
	
\begin{figure}[t]
		\centering
		\includegraphics[width=0.75\textwidth]{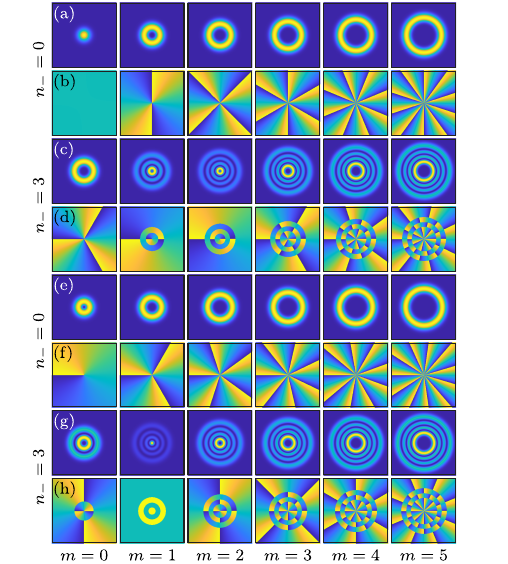}
		\caption{
			(a),(c),(e),(g) Irradiance and (b),(d),(f),(h) phase distributions for representative elements of the positive-circular-sector optical Bargmann basis.
			Rows show fixed $n_{-}$ values, $n_{-}=0$ and $n_{-}=3$, in the even, (a)-(d) $\kappa = 1/4$, and odd, (e)-(h) $\kappa = 3/4$, Bargmann sectors.
			Columns show increasing ladder index $m=0,1,2,\ldots$.
		}
		\label{fig:Fig1}
	\end{figure}

	For each circular mode, we define a single-mode $\mathrm{su}(1,1)$ algebra,
	\begin{align}
		\begin{aligned}
			\hat{K}_{+}^{(\pm)} =&~ \frac{1}{2} \hat{a}_{\pm}^{\dagger \, 2}, \\
			\hat{K}_{-}^{(\pm)} =&~ \frac{1}{2} \hat{a}_{\pm}^{2}, \\
			\hat{K}_{0}^{(\pm)} =&~ \frac{1}{2} \left( \hat{n}_{\pm} + \frac{1}{2} \right),
		\end{aligned}
	\end{align}
	with commutation relations
	\begin{align}
		\begin{aligned}
			\left[ \hat{K}_{0}^{(\pm)}, \hat{K}_{\pm}^{(\pm)} \right] =&~ \pm \hat{K}_{\pm}^{(\pm)}, \\
			\left[ \hat{K}_{-}^{(\pm)}, \hat{K}_{+}^{(\pm)} \right] =&~ 2 \hat{K}_{0}^{(\pm)}.
		\end{aligned}
	\end{align}
	The Casimir has the fixed value $\hat{C}^{(\pm)}=-3/16$, so the Bargmann index satisfies $\kappa \left( \kappa -1 \right)=-3/16$ and takes the values $\kappa=1/4,3/4$.
	The simultaneous eigenbasis of $\hat{n}_{\pm}$,
	\begin{align}
		\begin{aligned}
			\lvert n_{+}, n_{-} \rangle =&~ \lvert p,\ell \rangle, \\
			p =&~ \min\!\left( n_{+}, n_{-} \right), \\
			\ell =&~ n_{+} - n_{-},
		\end{aligned}
	\end{align}
	provides the natural representation basis for these algebras.
	\textcolor{black}{The circular occupation difference is the same azimuthal index $\ell$ introduced in the LGB profile.}
	The generators $\hat{K}_{\pm}^{(\pm)}$ change $n_{\pm}$ by two units and commute with $\hat{n}_{\mp}$.
	We use this action to parcel the LGB modal Hilbert space in two complementary ways.
	For the positive (negative) construction, $n_{-}$ ($n_{+}$) labels the fixed circular modal subspace and $n_{+}$ ($n_{-}$) runs along its modal basis ladder.
	Each parceling reproduces the full LGB modal Hilbert space, so the two parcelings are not independent labels to be used simultaneously.
	We write the corresponding basis vectors and running occupation number as
	\begin{align}
		\begin{aligned}
			\lvert n_{\mp}; \kappa, m \rangle_{\pm} =&~ \lvert n_{+}, n_{-} \rangle, \\
			n_{\pm} =&~ 2 \left( \kappa + m - \frac{1}{4} \right),
		\end{aligned}
	\end{align}
	where $m=0,1,2,\ldots$ counts the position along the selected ladder.
	For $\kappa=1/4$, the running occupation is even, $n_{\pm}=2m=0,2,4,\ldots$, while for $\kappa=3/4$, the running occupation is odd, $n_{\pm}=2m+1=1,3,5,\ldots$.
	The Bargmann index $\kappa$ selects the parity ladder, while $m$ counts the position along it.
	The generators act within each fixed-$n_{\mp}$ ladder,
	\begin{align}
		\begin{aligned}
			\hat{K}_{0}^{(\pm)} \lvert n_{\mp}; \kappa, m \rangle_{\pm} =&~ \left( \kappa + m \right) \lvert n_{\mp}; \kappa, m \rangle_{\pm}, \\
			\hat{K}_{+}^{(\pm)} \lvert n_{\mp}; \kappa, m \rangle_{\pm} =&~ \sqrt{ \left( m + 1 \right) \left( 2 \kappa + m \right)} \lvert n_{\mp}; \kappa, m + 1 \rangle_{\pm}, \\
			\hat{K}_{-}^{(\pm)} \lvert n_{\mp}; \kappa, m \rangle_{\pm} =&~ \sqrt{m \left( 2 \kappa + m - 1 \right)} \lvert n_{\mp}; \kappa, m - 1 \rangle_{\pm}.
		\end{aligned}
	\end{align}
	Figure~\ref{fig:Fig1} shows representative positive circular sector basis elements for $n_{-}=0,3$ across both Bargmann sectors and increasing ladder index $m$.
	In the even sector, $\kappa=1/4$, rows Fig.~\ref{fig:Fig1}(a) and Fig.~\ref{fig:Fig1}(c) show the irradiance distributions, while rows Fig.~\ref{fig:Fig1}(b) and Fig.~\ref{fig:Fig1}(d) show the corresponding phase distributions.
	In the odd sector, $\kappa=3/4$, rows Fig.~\ref{fig:Fig1}(e) and Fig.~\ref{fig:Fig1}(g) show the irradiance distributions, while rows Fig.~\ref{fig:Fig1}(f) and Fig.~\ref{fig:Fig1}(h) show the corresponding phase distributions.
	
	\section{Circular-mode squeezed LGBs}
	\label{sec:Sec3}
	
	We construct \textcolor{black}{generalized coherent state families of the $\mathrm{SU}(1,1)$ Lie group} on each fixed-$n_{\mp}$ ladder by applying the circular-mode squeezing operator
	\begin{align}
		\hat{S}^{(\pm)}(\zeta) =&~ e^{\zeta \hat{K}_{+}^{(\pm)} - \zeta^{\ast} \hat{K}_{-}^{(\pm)}}
	\end{align}
	to the optical Bargmann basis,
	\begin{align}
		\lvert n_{\mp}, \zeta; \kappa, m \rangle_{\pm} =&~ \hat{S}^{(\pm)}(\zeta) \lvert n_{\mp}; \kappa, m \rangle_{\pm},
	\end{align}
	where $\zeta = \lvert \zeta \rvert e^{i\theta}$ is the complex squeezing parameter.
	\textcolor{black}{For nonzero squeezing, $\theta = \arg \zeta$ is a real phase defined modulo $2\pi$, independent of $z$, that controls the rotation of the beam's intensity and phase profiles at the waist, $z=0$.}
	The seed $m = 0$ gives the standard $\mathrm{SU}(1,1)$ coherent state~\cite{Gilmore1972p391,Perelomov1972p222}, while seeds $m \geq 1$ give generalized $\mathrm{SU}(1,1)$ coherent states~\cite{VillanuevaVergara2015p22836,HuertaAlderete2019p2734}.
	The case $n_{\mp}=0$, $m=0$, and $\kappa=1/4$ gives the optical analogue of a quantum single-mode squeezed vacuum, while all other cases give optical analogues of quantum single-mode squeezed number states.
	\textcolor{r2revision}{Our squeezing operator implements a mathematical mode-space transformation of a classical optical field.}
	\textcolor{r2revision}{At fixed $\zeta$, the unitarity of $\hat{S}^{(\pm)}(\zeta)$ makes our squeezed basis orthonormal and complete in each subspace with fixed $n_{\mp}$ and $\kappa$ as $m=0,1,2,\ldots$ runs along the ladder. Including both parity sectors and all values of $n_{\mp}$ recovers the full LGB modal Hilbert space in either circular construction.}
	
	The squeezed states admit the expansion~\cite{VillanuevaVergara2015p22836,HuertaAlderete2019p2734,MoralesRodriguez2024p033523}
	\begin{align}
		\lvert n_{\mp},\zeta; \kappa, m \rangle_{\pm} =&~ \sum_{u=0}^{\infty} C_{u}^{(\kappa,m)}(\zeta) \lvert n_{\mp}; \kappa, u \rangle_{\pm},
	\end{align}
	with complex coefficients
	\begin{align}
		\begin{aligned}
			C_{u}^{(\kappa,m)}(\zeta) =&~ (-1)^{m} \mathrm{sech}^{2 \kappa} \lvert \zeta \rvert \sqrt{ \binom{2 \kappa + m - 1}{m} \binom{2 \kappa + u - 1}{u}} \\
			&~ \times \left( e^{-i \theta} \tanh \lvert \zeta \rvert \right)^{m} \left( e^{i \theta} \tanh \lvert \zeta \rvert \right)^{u} {}_{2}F_{1}\!\left( -u, -m; 2 \kappa; -\mathrm{csch}^{2}\lvert \zeta \rvert \right),
		\end{aligned}
	\end{align}
	where $\binom{a}{b} = \Gamma(a+1)/ ( \Gamma(b+1) \Gamma(a-b+1) )$ is the binomial coefficient expressed in terms of the gamma function, and ${}_{2}\mathrm{F}_{1}(a,b;c;z)$ is the Gauss hypergeometric function.
	
	We use this expansion to define our family of anisotropic squeezed LGBs as optical analogues of single-mode squeezed states,
	\begin{align}
		\begin{aligned}
			\Psi_{n_{\mp}, \zeta; \kappa, m}^{(\pm)}(\bm{\rho},z) =&~ \left\langle \bm{\rho},z \middle\vert n_{\mp}, \zeta; \kappa, m \right\rangle_{\pm}, \\
			=&~ \sum_{u=0}^{\infty} C_{u}^{(\kappa,m)}(\zeta) \Psi_{p(u), \ell_{\pm}(u)}(\bm{\rho},z).
		\end{aligned}
	\end{align}
	\textcolor{black}{Adjacent ladder components differ by two in total modal order, so propagation multiplies the squeezing phase factor $e^{i\theta}$ by $e^{2i\varphi(z)}$, giving the phase $\theta+2\varphi(z)$.}
	
\begin{figure}[t]
		\centering
		\includegraphics[width=0.75 \textwidth]{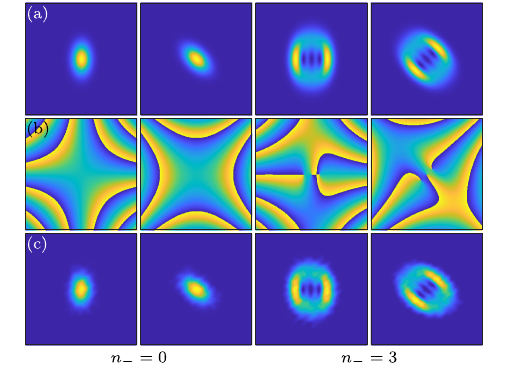}
		\caption{
			(a) Irradiance $\lvert \Psi_{n_{-}, \zeta; \kappa, 0}^{(+)}(q_{x}, q_{y}, q_{z}) \rvert^{2}$ and (b) phase $\operatorname{arg} \Psi_{n_{-}, \zeta; \kappa, 0}^{(+)}(q_{x}, q_{y}, q_{z})$ for representative standard coherent states in the positive circular sector with $\kappa = 1/4$.
			Columns show, from left to right, $n_{-}=0$ with $\{ \lvert \zeta \rvert, \theta \} = \{ 1/2, \pi \}$ and $\{ \lvert \zeta \rvert, \theta \} = \{ 1/2, \pi/2 \}$, and $n_{-}=3$ with the same real and complex squeezings. (c) Experimental intensity distributions corresponding to the patterns shown in (a). \textcolor{black}{For the experimental results, we set $w_0 = 0.24$ $mm$ and $\lambda = 632.8$ $nm$.}
		}
		\label{fig:Fig2}
	\end{figure}

	For the lowest-weight seed, $m=0$, our anisotropic squeezed LGBs factor into the Gaussian beam curvature and Gouy phases, a parity-dependent polynomial, and a squeezed Gaussian envelope,
	\begin{align}
		\begin{aligned}
			\Psi_{n_{\mp}, \zeta; \kappa, 0}^{(\pm)}(q_{x}, q_{y}; z) =&~ \sqrt{ \frac{2}{\pi n_{\mp}!} } \, \frac{\mathrm{sech}^{2\kappa}\lvert \zeta \rvert}{w(z)}
			e^{ - i \frac{k w^{2}(z)}{4 R(z)} \left( q_{x}^{2} + q_{y}^{2} \right)}
			e^{i \left( n_{\mp} + 2 \kappa + \frac{1}{2} \right) \varphi(z)}  \\
			&~ \times \mathcal{P}_{\kappa,n_{\mp}}^{(\pm)}(q_{x}, q_{y}, q_{z}) \Phi_{\pm}(q_{x}, q_{y}, q_{z}),
		\end{aligned}
	\end{align}
	where the polynomial factor,
	\begin{align}
		\mathcal{P}_{\kappa,n_{\mp}}^{(\pm)}(q_{x},q_{y},q_{z}) =&~
		\begin{cases}
			H_{n_{\mp}} \! \left( X_{\pm}, \frac{1}{2} \eta(z) \right), & \kappa = \frac{1}{4}, \\
			\left( q_{x} \pm i q_{y} \right) H_{n_{\mp}}\!\left( X_{\pm}, \frac{1}{2} \eta(z) \right) - n_{\mp} H_{n_{\mp} - 1}\!\left( X_{\pm}, \frac{1}{2} \eta(z) \right), & \kappa = \frac{3}{4},
		\end{cases}
	\end{align}
	depends on the Gouy-dressed squeezing parameter,
	\begin{align}
		\begin{aligned}
			\eta(z) =&~ \tanh \lvert \zeta \rvert \, e^{i\Theta(z)}, \\
			\Theta(z) =&~ 2 \varphi(z) + \theta ,
		\end{aligned}
	\end{align}
	through the argument
	\begin{align}
		X_{\pm}(q_{x},q_{y},q_{z}) =&~ q_{x} \mp i q_{y} - \eta(z) \left( q_{x} \pm i q_{y} \right).
	\end{align}
	Here, $H_{n}(x,y) = n! \sum_{s=0}^{\lfloor n/2\rfloor} x^{n-2s} y^{s} / \left[ \left( n-2s \right)! s! \right]$ is the two-variable Hermite polynomial, and the squeezed Gaussian envelope is
	\begin{align}
		\Phi_{\pm}(q_{x},q_{y},q_{z}) =&~ e^{-\frac{1}{2}\left( q_{x}^{2} + q_{y}^{2} \right) + \frac{1}{2} \eta(z) \left( q_{x} \pm i q_{y} \right)^{2}}.
	\end{align}
	The amplitude of the Gouy-dressed squeezing parameter, $\lvert \eta(z) \rvert = \tanh \lvert \zeta \rvert$, fixes the squeezing strength, while its phase, $\arg \eta(z) = 2 \varphi(z) + \theta$, sets the transverse orientation of the anisotropic Gaussian envelope.
	Figure~\ref{fig:Fig2}(a) shows the lowest-weight positive-circular-sector irradiance at the waist plane for $n_{-}=0,3$, $\lvert \zeta \rvert=1/2$, and $\theta=\pi,\pi/2$, and Fig.~\ref{fig:Fig2}(b) shows the corresponding phase profiles.
	Changing the squeezing phase rotates the irradiance and phase profiles without changing their scaled transverse shape.
	The experimental irradiance patterns in Fig.~\ref{fig:Fig2}(c) confirm the rotation.
	
\begin{figure}[t]
		\centering
		\includegraphics[width=0.75\textwidth]{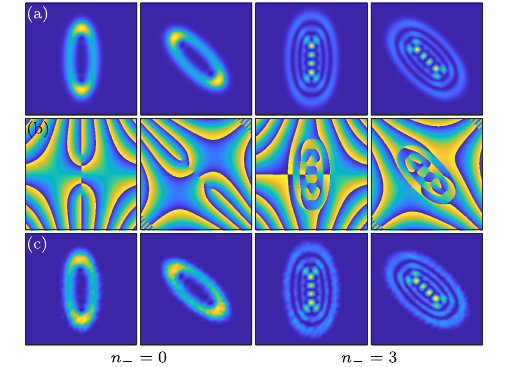}
		\caption{
			Representative generalized coherent states in the positive circular mode with $\kappa = 1/4$. The rows depict (a) the theoretical irradiance, $\lvert \Psi_{n_{-}, \zeta; \kappa, m}^{(+)}(q_{x}, q_{y}, q_{z}) \rvert^{2}$, (b) the corresponding phase, $\operatorname{arg} \Psi_{n_{-}, \zeta; \kappa, m}^{(+)}(q_{x}, q_{y}, q_{z})$, and \textcolor{black}{(c) experimental intensity distributions corresponding to the patterns shown in (a)}. 
			From left to right, the columns correspond to $n_{-}=0$ with $m=3$ and $\{ \lvert \zeta \rvert, \theta \} = \{ 1/2, \pi \}$, $n_{-}=0$ with $m=3$ and $\{ \lvert \zeta \rvert, \theta \} = \{ 1/2, \pi/2 \}$, $n_{-}=3$ with $m=3$ and $\{ \lvert \zeta \rvert, \theta \} = \{ 1/2, \pi \}$, and $n_{-}=3$ with $m=3$ and $\{ \lvert \zeta \rvert, \theta \} = \{ 1/2, \pi / 2 \}$. \textcolor{black}{For the experimental results, we set $w_0 = 0.24$ $mm$ and $\lambda = 632.8$ $nm$.}
		}
		\label{fig:Fig3}
	\end{figure}

	For excited seeds, $m \geq 1$, we built our anisotropic squeezed LGBs from the lowest-weight form by recursion.
	The Bargmann basis satisfies
	\begin{align}
		\lvert n_{\mp}; \kappa, m \rangle_{\pm} =&~ \frac{1}{\sqrt{m! \left( 2 \kappa \right)_{m}}} \left( \hat{K}_{+}^{(\pm)} \right)^{m} \lvert n_{\mp}; \kappa, 0 \rangle_{\pm},
	\end{align}
	where $\left( 2 \kappa \right)_{m} = \Gamma\left( 2 \kappa + m \right) / \Gamma\left( 2 \kappa \right)$ is the Pochhammer symbol.
	Conjugating by the squeezing operator gives
	\begin{align}
		\lvert n_{\mp}, \zeta; \kappa, m \rangle_{\pm} =&~ \frac{1}{\sqrt{m! \left( 2 \kappa \right)_{m}}} \left( \hat{\mathcal{K}}_{+}^{(\pm)} \right)^{m} \lvert n_{\mp}, \zeta; \kappa, 0 \rangle_{\pm},
	\end{align}
	with transformed generator
	\begin{align}
		\begin{aligned}
			\hat{\mathcal{K}}_{+}^{(\pm)} =&~ \hat{S}^{(\pm)}(\zeta) \hat{K}_{+}^{(\pm)} \hat{S}^{(\pm)\dagger}(\zeta), \\
			=&~ \cosh^{2}\lvert \zeta \rvert \left[
			\hat{K}_{+}^{(\pm)}
			+ e^{-i2\theta}\tanh^{2}\lvert \zeta \rvert \, \hat{K}_{-}^{(\pm)}
			- 2 e^{-i\theta}\tanh\lvert \zeta \rvert \, \hat{K}_{0}^{(\pm)}
			\right].
		\end{aligned}
	\end{align}
	The factorization $\left( 2 \kappa \right)_{m} = \left( 2 \kappa \right)_{m-1}\left( 2 \kappa + m - 1 \right)$ yields the optical recursion \textcolor{r2revision}{at the waist,}
	\begingroup\color{r2revision}
	\begin{align}
		\Psi_{n_{\mp}, \zeta; \kappa, m}^{(\pm)}(q_{x},q_{y},0) =&~ \frac{1}{\sqrt{m \left( 2 \kappa + m - 1 \right)}} \, \hat{\mathcal{D}}_{\pm}(\zeta) \Psi_{n_{\mp}, \zeta; \kappa, m - 1}^{(\pm)}(q_{x},q_{y},0),
	\end{align}
	\endgroup
	where the differential operator,
	\begin{align}
		\begin{aligned}
			\hat{\mathcal{D}}_{\pm}(\zeta) =&~ \frac{1}{8}\cosh^{2}\lvert \zeta \rvert
			\left\{
			\left[ \bm{e}_{\pm} \cdot \left( \bm{q} - \bm{\nabla} \right) \right]^{2}
			+ e^{-i2\theta}\tanh^{2}\lvert \zeta \rvert \left[ \bm{e}_{\mp} \cdot \left( \bm{q} + \bm{\nabla} \right) \right]^{2} \right. \\
			&~ \left.
			- 2e^{-i\theta}\tanh\lvert \zeta \rvert \left( q_{\rho}^{2} - \nabla^{2} \pm 2 \hat{L}_{z} \right)
			\right\},
		\end{aligned}
	\end{align}
	realizes $\hat{\mathcal{K}}_{+}^{(\pm)}$ on the dimensionless transverse profile.
	Here, $\bm{e}_{\pm} = \left( 1, \pm i \right)^{T}$ are the circular basis vectors, $\bm{q} = \left( q_{x}, q_{y} \right)^{T}$ is the dimensionless transverse position vector, $\bm{\nabla} = \left( \partial_{q_{x}}, \partial_{q_{y}} \right)^{T}$ is the transverse gradient, and $\hat{L}_{z} = - i \left( q_{x} \partial_{q_{y}} - q_{y} \partial_{q_{x}} \right)$ is the longitudinal orbital angular momentum operator.
	Excited seeds generate polynomial deformations of the same squeezed Gaussian envelope $\Phi_{\pm}$, with the polynomial degree increasing by two at each step in $m$.
	\textcolor{r2revision}{We propagate these waist fields through their LGB modal expansion.}
	Figure~\ref{fig:Fig3}(a) shows representative generalized-coherent-state irradiance profiles generated from the third excited Bargmann seed, $m=3$, in the positive circular sector for $n_{-}=0,3$, $\lvert \zeta \rvert=1/2$, and $\theta=\pi,\pi/2$, and Fig.~\ref{fig:Fig3}(b) shows the corresponding phase profiles.
	For fixed $n_{-}=0,3$, the excited seed adds internal nodal structure to the squeezed envelope.
	Changing the squeezing phase rotates the irradiance and phase profiles without changing their scaled transverse shape.
	The experimental irradiance patterns in Fig.~\ref{fig:Fig3}(c) confirm the rotation.
	
	\section{\texorpdfstring{\textcolor[RGB]{0,0,0}{Gouy phase driven self-similar rigid rotation}}{Gouy phase driven self-similar rigid rotation}}
	\label{sec:Sec4}
	
	\begingroup\color{r2revision}
	In each subspace with fixed circular occupation number $n_{\mp}$, the total transverse modal number $N$ and azimuthal index $\ell$ satisfy $N=\pm\ell+2n_{\mp}$, so the modal Gouy phase factor separates as $e^{i(N+1)\varphi(z)}=e^{i(2n_{\mp}+1)\varphi(z)}e^{\pm i\ell\varphi(z)}$.
	The common phase and angular shift give the full-field propagation law in physical polar coordinates,
	\begin{align}
		\begin{aligned}
			\Psi_{n_{\mp}, \zeta; \kappa, m}^{(\pm)}(\rho,\phi;z) =&~ \frac{w_{0}}{w(z)} e^{-i\frac{k\rho^{2}}{2R(z)}} e^{i\left(2n_{\mp}+1\right)\varphi(z)}  \Psi_{n_{\mp}, \zeta; \kappa, m}^{(\pm)}\!\left(\frac{w_{0}}{w(z)}\rho,\phi\pm\varphi(z);0\right).
		\end{aligned}
		\label{eq:full-field-rotation}
	\end{align}
	For every ladder index $m$, this identity establishes the self-similar rigid rotation by $\mp\varphi(z)$ of the full intensity and nodal pattern.
	Factoring out the curvature phase reveals the same self-similar rigid rotation of the transverse phase profile, up to the spatially uniform Gouy phase $\left(2n_{\mp}+1\right)\varphi(z)$.
	Exchanging the circular occupation numbers maps corresponding fields between the Hilbert subspace constructions labeled by the fixed circular occupation numbers $n_{\mp}$ through the azimuthal reflection $\phi\mapsto-\phi$, reversing the rotation handedness.
	\endgroup
	
	The Gouy-dressed squeezing parameter, $\eta(z)$, controls our anisotropic Gaussian envelope.
	For fixed squeezing amplitude $\lvert \zeta \rvert$, propagation changes only $\Theta(z)$.
	The envelope satisfies
	\begin{align}
		\begin{aligned}
			\Phi_{\pm}(q_{x},q_{y},q_{z}) =&~ e^{-\frac{1}{2} \left( q_{x}^{2} + q_{y}^{2} \right) + \frac{1}{2} \lvert \eta \rvert e^{i\Theta(z)} \left( q_{x} \pm i q_{y} \right)^{2}} \\
			=&~ e^{-\frac{1}{2} \left[ 1 - \lvert \eta \rvert e^{i\Theta(z)} \right] q_{x}^{2}
				-\frac{1}{2} \left[ 1 + \lvert \eta \rvert e^{i\Theta(z)} \right] q_{y}^{2}
				\pm i \lvert \eta \rvert e^{i\Theta(z)} q_{x}q_{y}}.
		\end{aligned}
	\end{align}
	Its irradiance is the real anisotropic Gaussian
	\begin{align}
		\left\lvert \Phi_{\pm}(q_{x},q_{y},q_{z}) \right\rvert^{2} =&~ e^{-\bm{q}^{T}\bm{G}_{\pm}(z)\bm{q}},
	\end{align}
	with shape matrix
	\begin{align}
		\bm{G}_{\pm}(z) =&~
		\begin{pmatrix}
			G_{xx}^{(\pm)}(z) & G_{xy}^{(\pm)}(z) \\
			G_{yx}^{(\pm)}(z) & G_{yy}^{(\pm)}(z)
		\end{pmatrix},
	\end{align}
	where $G_{xx}^{(\pm)}(z) = 1 - \lvert \eta \rvert \cos \Theta(z)$, $G_{xy}^{(\pm)}(z) = G_{yx}^{(\pm)}(z) = \pm \lvert \eta \rvert \sin \Theta(z)$ and $G_{yy}^{(\pm)}(z) = 1 + \lvert \eta \rvert \cos \Theta(z)$.
	The eigenvalues $g_{\mathrm{min}} = 1 - \lvert \eta \rvert$ and $g_{\mathrm{max}} = 1 + \lvert \eta \rvert$ do not depend on $z$, so propagation preserves the scaled semi-axis widths,
	\begin{align}
		\begin{aligned}
			\sigma_{\mathrm{maj}} =&~ \frac{1}{\sqrt{g_{\mathrm{min}}}}, \\
			\sigma_{\mathrm{min}} =&~ \frac{1}{\sqrt{g_{\mathrm{max}}}},
		\end{aligned}
	\end{align}
	and the aspect ratio $\sigma_{\mathrm{maj}}/\sigma_{\mathrm{min}} = \sqrt{g_{\mathrm{max}}}/ \sqrt{g_{\mathrm{min}}}$.
	Propagation preserves the scaled transverse shape and changes only the inclination angle of the principal axes.
	The major-axis angle $\beta_{\pm}(z)$ with respect to the $q_{x}$-axis satisfies
	\begin{align}
		\tan 2\beta_{\pm}(z) =&~ \frac{2G_{xy}^{(\pm)}(z)}{G_{xx}^{(\pm)}(z)-G_{yy}^{(\pm)}(z)}
		=~ \mp \tan \Theta(z),
	\end{align}
	which gives the orientation angle,
	\begin{align}
		\beta_{\pm}(z) =&~ \mp \frac{1}{2}\Theta(z) \mod \pi 
		=~ \mp \varphi(z) \mp \frac{\theta}{2}  \mod \pi.
		\label{eq:beta}
	\end{align}
	Propagation from $z_{0}$ to $z_{1}$ rotates the major axis by the accumulated Gouy phase,
	\begin{align}
		\beta_{\pm}(z_{1}) - \beta_{\pm}(z_{0}) =&~ \mp \left[ \varphi(z_{1}) - \varphi(z_{0}) \right] \mod \pi.
	\end{align}
	The Rayleigh range gives a finite propagation interval with a large rotation angle.
	At the waist plane, $\beta_{\pm}(0) = \mp \theta /2 \mod \pi$.
	At the Rayleigh planes, $\varphi(s z_{R}) = s\pi/4$ with $s = \pm1$, so $\beta_{\pm}(s z_{R}) = \beta_{\pm}(0) \mp s\pi/4 \mod \pi$.
	Propagation from $z=-z_{R}$ to $z=+z_{R}$ rotates the principal axis by $\mp \pi/2$ while preserving the scaled semi-axis widths and their ratio.
	\textcolor{black}{The rotation angle over this interval is independent of the modal indices. Bessel-mode superpositions likewise convert relative propagation phase into rotation and can produce vortex braiding through multiple turns~\cite{Voitiv2020p1321}, while ring-Airy superpositions can produce accelerating rotation~\cite{Brimis2020p280}.}
	
	\begingroup\color{r2revision}
	From our orientation angle, we obtain the angular sensitivity on a continuous branch,
	\begin{align}
		\frac{d\beta_{\pm}}{dz} =&~ \mp \frac{1}{z_{R}\left(1+q_{z}^{2}\right)},
		\label{eq:angular-sensitivity}
	\end{align}
	and, for a known waist position, reference orientation, and Rayleigh range, the local axial uncertainty through first-order propagation of small angular fluctuations~\cite{JCGM2008},
	\begin{align}
		\sigma_{z} \simeq&~ \left\lvert \frac{d\beta_{\pm}}{dz} \right\rvert^{-1}\sigma_{\beta}
		= z_{R}\left(1+q_{z}^{2}\right)\sigma_{\beta},
		\label{eq:axial-uncertainty}
	\end{align}
	where the angular uncertainty $\sigma_{\beta}$ is the standard deviation, in radians, of orientation estimates at a fixed propagation plane.
	The angular uncertainty can be estimated from the sample standard deviation of orientation estimates extracted from repeated intensity images on a common angular branch, following the repeatability approach used, for example, in rotating point-spread-function microscopy~\cite{Thompson2010p211}.
	The angular sensitivity decreases in magnitude with distance from the waist plane $z=0$, so the local axial uncertainty increases for fixed angular uncertainty.
	\endgroup
	
\begin{figure}[t]
		\centering
		\includegraphics[width=0.75\textwidth]{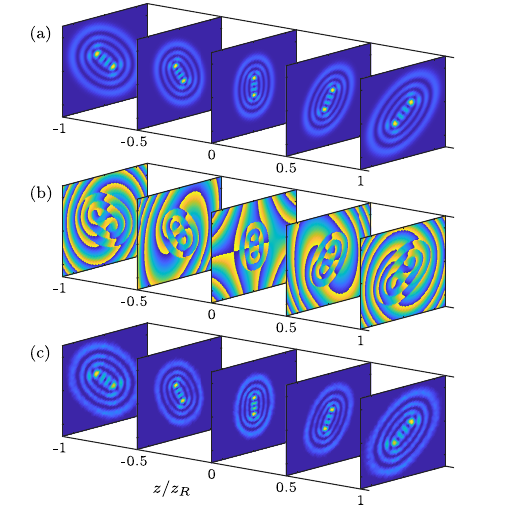}
		\caption{
			Self-similar rigid rotation of an anisotropic squeezed LGB from $z=-z_{R}$ to $z=z_{R}$ in steps of $z_{R}/2$.
			(a) Theoretical irradiance for $n_{-}=3$, $m=3$, $\kappa=1/4$, and $\{ \lvert \zeta \rvert,\theta \}=\{1/3,-\pi\}$.
			Each irradiance distribution is independently normalized at its propagation plane.
			(b) Corresponding phase distributions.
			(c) Experimental irradiance distributions corresponding to (a). \textcolor{black}{For the experimental results, we set $w_0 = 0.24$ $mm$, $z_R=28.5$ $cm$, and $\lambda = 632.8$ $nm$.}
		}
		\label{fig:Fig4}
	\end{figure}

	Figure~\ref{fig:Fig4}(a) shows the self-similar rigid rotation of the positive-circular-sector irradiance for $n_{-}=3$, $m=3$, $\kappa=1/4$, and $\{ \lvert \zeta \rvert,\theta \}=\{1/3,-\pi\}$ over the interval $-z_{R}\leq z \leq z_{R}$ in steps of $z_{R}/2$.
	Each irradiance distribution is independently normalized at its propagation plane.
	Figure~\ref{fig:Fig4}(b) shows the corresponding phase profiles, where \textcolor{r2revision}{the radial curvature phase produces the spiral structure away from the waist.}
	The experimental irradiance patterns in Fig.~\ref{fig:Fig4}(c) confirm the predicted \textcolor{r2revision}{self-similar rigid rotation}.
	
\begin{figure}[t]
		\centering
		\includegraphics[width=0.75\textwidth]{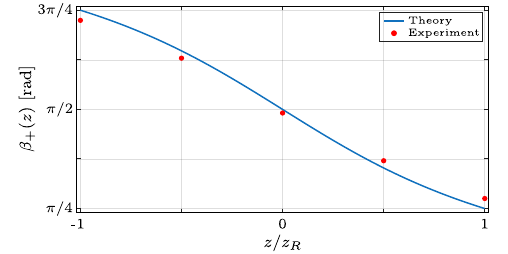}
		\caption{Theoretical inclination angle $\beta_{+}(z)$ (solid blue line) and experimental angle $\beta_{exp}$ (red circles) as \textcolor{black}{a} function of the normalized propagation distance $q_{z} = z / z_{R}$. 
			The experimental angles were obtained from the second-moment analysis of the intensity distributions shown in Figure~\ref{fig:Fig4}(c).}
		\label{fig:Fig5}
	\end{figure}

	To compare the experimental and theoretical rotation angles, we determined the inclination angle, $\beta_{\mathrm{exp}}$, of the \textcolor{black}{major} principal axis of each experimental intensity distribution using the second-moment beam characterization method~\cite{ISO11146-3,Gao2005},
	\begin{align}\label{eq:betaexp}
		\textcolor{black}{\beta_{\text{exp}} = \frac{1}{2} \operatorname{atan2}\left(2M_{xy}, M_{xx}-M_{yy}\right)},
	\end{align}
	\textcolor{black}{where $\operatorname{atan2}(a,b)$ is the two-argument arctangent that determines the angular quadrant from the signs of both arguments.}
	\textcolor{black}{The normalized second central moments are}
	\begin{align}
		\begin{aligned}
			M_{xx} =&~ \frac{1}{P} \iint_{-\infty}^{\infty} dx dy \, x_{c}^{2} I(x,y) , \\
			M_{yy} =&~ \frac{1}{P} \iint_{-\infty}^{\infty} dx dy \, y_{c}^{2} I(x,y) , \\
			M_{xy} =&~ \frac{1}{P} \iint_{-\infty}^{\infty} dx dy \, x_{c} y_{c} I(x,y) ,
		\end{aligned}
	\end{align}
	where $P=\iint I(x,y)\,dx\,dy$ is the total power, $I(x,y)$ is the intensity recorded by the CCD at the propagation distance $z$, and $x_{c} = x - \bar{x}$ and $y_{c} = y -\bar{y}$ are coordinates measured with respect to the intensity-weighted beam centroid,
	\begin{align}
		\begin{aligned}
			\bar{x} =&~ \frac{1}{P} \iint_{-\infty}^{\infty} dx dy \, x I(x,y) , \\
			\bar{y} =&~ \frac{1}{P} \iint_{-\infty}^{\infty} dx dy \, y I(x,y) .
		\end{aligned}
	\end{align}
	\noindent \textcolor{black}{Since the principal-axis orientation is defined modulo $\pi$, equivalent angular values differing by integer multiples of $\pi$ were selected to follow the same continuous angular branch as the theoretical prediction. The deviation between the experimental and theoretical angles at each experimental position was expressed as 
		\begin{equation}
			\Delta \beta(z_{j}) = \beta_{\text{exp}}(z_{j})-\beta_{+}(z_{j}).
		\end{equation}
		To quantify the overall agreement between experiment and theory, we use the root-mean-square deviation ($\mathrm{RMSD}$),
		\begin{equation}
			\mathrm{RMSD} = \sqrt{\frac{1}{N}\sum_{j=1}^{N}\left[\Delta\beta(z_{j})\right]^{2}}, \;\;\;\;\; N = 5,
		\end{equation}
		which yielded $0.0635~\mathrm{rad}$ ($3.64^{\circ}$) for the experimental positions $(z_{j})$, with $j = 1, \ldots, 5$, shown in Fig. 5.}
	\textcolor{r2revision}{Our experiment is a single-shot proof-of-principle demonstration, with one intensity image acquired at each propagation position. The RMSD quantifies the deviation of the measured orientations from the predicted angular dependence of the self-similar rigid rotation.}
	Figure~\ref{fig:Fig5} compares the theoretical inclination angle $\beta_{+}(z)$ (solid line) with the experimental angle $\beta_{exp}$ (red circles), obtained from the second-moment analysis of the experimental intensity distributions shown in Figure~\ref{fig:Fig4}(c). 
	\textcolor{black}{The experimental results follow the predicted \textcolor{r2revision}{self-similar rigid rotation}, with an RMSD of $3.64^{\circ}$ over the five measured positions.} 
	
	\section{Generation of anisotropic squeezed LGBs using synthetic phase holograms}
	\label{sec:Sec5}
	
	We generate our anisotropic squeezed LGBs with synthetic phase holograms (SPHs)~\cite{Arrizon2007p3500,Mellado2021p1094}.
	At a chosen reference plane $z = z_{0}$, the target normalized complex amplitude has the form
	\begin{align}
		f(x,y; z_{0}) =&~ a(x,y; z_{0}) e^{i \phi(x,y;z_{0})},
		\label{eq:ca}
	\end{align}
	where $0 \leq a(x,y; z_{0}) \leq 1$ and $-\pi \leq \phi(x,y; z_{0}) \leq \pi$.
	For the measurements in Figs.~\ref{fig:Fig2} and \ref{fig:Fig3}, the reference plane is the beam waist, $z_{0}=0$.
	For the propagation measurements in Fig.~\ref{fig:Fig4}, the reference plane is the initial propagation plane, $z_{0}=-z_{R}$.
	
	The transmittance function of an SPH that encodes the target beam can be written as
	\begin{align}
		h(x,y;z_{0}) =&~ e^{ i \mathcal{E}[a(x,y;z_{0})] \sin[\phi(x,y;z_{0})+2\pi u_{0}(x+y)]},
		\label{eq:sph}
	\end{align}
	where $u_{0}$ is the carrier frequency. \textcolor{black}{Using the Jacobi-Anger expansion~\citep{Arfken2005p687}, the transmittance function can be expressed as
		\begin{align}
			h(x,y;z_{0}) =&~ \sum_{q=-\infty}^{\infty} h_q(x,y;z_0) e^{i2\pi qu_{0}(x+y)},
			\label{eq:sph_expansion}
		\end{align}
		where $h_q(x,y;z_0)=J_q\{\mathcal{E}[a(x,y;z_0)]\} e^{iq\phi(x,y;z_0)}$ is the complex amplitude of the $q$-th diffraction order of the SPH, and $J_q$ is the $q$-th Bessel function of the first kind.} The target field is encoded in the first diffraction order ($q=1$) when
	\begin{align}
		J_{1}\!\left\{\mathcal{E}[a(x,y;z_{0})] \right\} =&~ \mathcal{C} a(x,y;z_{0}),
	\end{align}
	where $\mathcal{C} \simeq 0.5819$ is the maximum value attained by $J_{1}$. Thus, the modulation function $\mathcal{E}[a(x,y;z_{0})]$ is determined by inverting the preceding relation.
	
	\textcolor{r2revision}{We obtained the reconstructed complex field $f^{\prime}$ numerically from the SPH using two fast Fourier transforms with spatial filtering in the intermediate Fourier plane. To quantify the agreement between the target complex field $f$ and the reconstructed complex field $f^{\prime}$, we calculate the root-mean-square (rms) error,}  
	\begin{equation}
		\mathrm{RMS} = \left[\frac{\iint_{\Omega} d\Omega \,  \lvert f - \gamma \cdot e^{i\alpha} \cdot f^{\prime} \rvert^{2}}{\iint_{\Omega} \; \lvert f \rvert^{2} \; d\Omega} \right]^{1/2},
	\end{equation}
	where $\Omega$ is the integration domain, while $\gamma$ and $\alpha$ are constants chosen to provide the best fit between the fields. 
	The RMS error \textcolor{black}{vanishes} for identical fields and increases toward unity as their difference grows. 
	For the examples shown in Figs.~\ref{fig:Fig2}--\ref{fig:Fig4}, it remains within the range $[0.04,\,0.05]$, \textcolor{r2revision}{indicating accurate numerical reconstruction of the target anisotropic squeezed LGBs.}
	
\begin{figure}[t]
		\centering
		\includegraphics[width=0.75\textwidth]{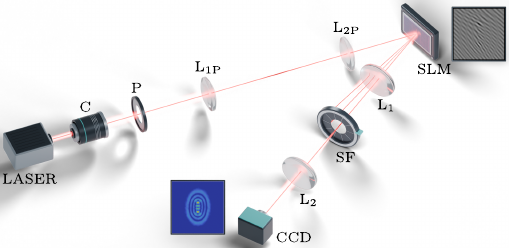}
		\caption{
			Experimental setup.
			Linearly polarized laser source; C,  collimator;
			P, circular pupil;
			$\mathrm{L}_{\mathrm{1P}}$, $\mathrm{L}_{\mathrm{2P}}$, $\mathrm{L}_{1}$, and $\mathrm{L}_{2}$, lenses;
			SLM, spatial light modulator;
			SF, spatial filter;
			CCD, camera. Schematic generated using \textit{Blender Optical Schematics}~\cite{rodriguez_lara_2026_optical_schematics}
		}
		\label{fig:Fig6}
	\end{figure}

	Figure~\ref{fig:Fig6} shows the experimental setup.
	We employ a collimator (C) to produce
    an expanded and collimated linearly polarized He--Ne laser beam (LB) at wavelength $\lambda=632.8~\mathrm{nm}$. This light illuminates a circular pupil (P).
	A double Fourier-transform system formed by lenses $\mathrm{L}_{\mathrm{1P}}$ and $\mathrm{L}_{\mathrm{2P}}$ images the pupil onto the spatial light modulator (SLM, Holoeye PLUTO VIS), where we implement the SPHs.
	Lens $\mathrm{L}_{1}$ Fourier transforms the SPH field onto the spatial-filter (SF) plane.
	The SPH produces several diffraction orders, and the SF selects the $+1$ order carrying the encoded anisotropic squeezed LGB.
	Lens $\mathrm{L}_{2}$ Fourier transforms the filtered field onto the output plane.
	We record the irradiance patterns in Figs.~\ref{fig:Fig2}--\ref{fig:Fig4} with a CCD camera (Hamamatsu, Model XC-77).
	For the propagation measurements in Fig.~\ref{fig:Fig4}, we translate the CCD along the optical axis with respect to the output plane.
	\textcolor{black}{We measured only intensity; the displayed phase profiles are theoretical. Phase reconstruction with an SLM-generated collinear reference is possible through phase-shifting holography~\cite{Andersen2019p404}, but lies outside the scope of this work.}
	
	\section{Conclusion}
	\label{sec:Sec6}
	
	We introduced an analytically complete family of anisotropic squeezed Laguerre--Gaussian beams.
	Our construction organizes the LGB modal Hilbert space into circular modal subspaces carrying parity-resolved single-mode $\mathrm{su}(1,1)$ representations and applies the corresponding squeezing operators to the resulting optical Bargmann bases.
	One circular occupation number labels each modal subspace, the other locates the state along the modal basis ladder, and the Bargmann index identifies the parity of the running occupation number.
	These indices determine the LGB modal components, modal order, and nodal structure.
	The complex squeezing parameter controls the transverse anisotropy and orientation of the squeezed Gaussian envelope.
	
	For the lowest-weight seed, the optical analogue of a standard $\mathrm{SU}(1,1)$ coherent state, we obtained a closed optical form that factors into Gaussian beam phases, a parity-dependent polynomial, and a squeezed Gaussian envelope.
	The polynomial factor contains two-variable Hermite polynomials, and both the polynomial factor and the squeezed Gaussian envelope depend on the same Gouy-dressed squeezing parameter.
	For excited seeds, the optical analogues of generalized $\mathrm{SU}(1,1)$ coherent states, we derived a recursion formula that builds the full family from the lowest-weight form by repeated application of the transformed raising generator.
	Each step increases the polynomial degree by two while preserving the same squeezed Gaussian envelope.
	
	The Gouy-dressed squeezing parameter controls the \textcolor{r2revision}{self-similar rigid rotation} of the anisotropic squeezed LGB family.
	For fixed squeezing amplitude, propagation changes only its phase, leaving the scaled semi-axis widths and aspect ratio invariant at every propagation plane.
	The principal-axis angle changes by the accumulated Gouy phase, and the two circular constructions rotate with opposite handedness.
	Our experimental results confirm the predicted self-similar rigid rotation.
	
	Our beam family provides a structured-light platform with tunable transverse anisotropy and Gouy-phase-controlled orientation.
	This orientation encodes longitudinal displacement and may support range finding, axial localization, propagation-sensitive alignment, optical manipulation, and modal diagnostics.
	Superpositions across circular modal subspaces may generate richer nodal structures while retaining Gouy-phase-controlled rotation.
	Vector-beam extensions may produce polarization textures with propagation-controlled rigid rotation.
    
	

	\section*{Acknowledgments}
	B.~M.~R.~L. thanks Jacinta Alderete Galan for providing daycare support throughout this work.
	He also acknowledges support and hospitality as an affiliate visiting colleague at the Department of Physics and Astronomy, University of New Mexico. D.~A.~O. acknowledges the Instituto Nacional de Astrof\'isica, \'Optica y Electr\'onica (INAOE) for its support as an affiliate researcher. G.~M.~V. acknowledges support from a postdoctoral fellowship awarded by the Secretar\'ia de Ciencia, Humanidades, Tecnolog\'ia e Innovaci\'on (SECIHTI), Mexico.
	
	\section*{Disclosures}
	The authors declare no conflicts of interest.
	
	\section*{Data Availability Statement}
	The data that support the findings of this study are available from the corresponding author upon reasonable request.
	
	
	\makeatletter
	\let\revision@bibitem\bibitem
	\renewcommand{\bibitem}[2][]{%
		\in@{#2}{Lusk2023p023509,Voitiv2024p016135,Lin2024p6178,Huang2016p132,Stoler1981p334,vanEnk1992p147,Voitiv2020p1321,Brimis2020p280,Andersen2019p404,Arfken2005p687}%
		\ifin@\color{black}\else\color{black}\fi
		\in@{#2}{Schechner1996pR50,JCGM2008,Thompson2010p211}%
		\ifin@\color{r2revision}\fi
		\revision@bibitem[#1]{#2}%
	}
	\makeatother

\end{document}